# INTERPRETING THE TILT-AND-TORSION METHOD TO EXPRESS SHOULDER JOINT KINEMATICS

Félix Chénier[1,2]*, Ilona Alberca[3], Arnaud Faupin[3], Dany H Gagnon[2,4]

[1]Mobility and Adaptive Sports Research Lab, Department of Physical Activity Sciences, Faculty of Sciences, Université du Québec à Montréal, Montreal, Canada

[2]Centre for Interdisciplinary Research in Rehabilitation of Greater Montreal (CRIR), Montreal, Canada

[3]Université de Toulon, Impact de l'Activité Physique sur la Santé (UR IAPS n°201723207F), Campus de La Garde, CS60584, F-83041 Toulon, France

[4]School of Rehabilitation, Faculty of Medicine, Université de Montréal, Montreal, Canada

*Corresponding author:

Félix Chénier, PhD
Associate professor, Department of Physical Activity Science, UQAM
chenier.felix@uqam.ca
+1-514-987-3000, ext. 5553

Postal address:

Félix Chénier
Biological Sciences Building, Office SB-4455
Université du Québec à Montréal
P.O. Box 8888, Station Centreville
Montreal, Quebec H3C 3P8 Canada

# ABSTRACT

*Background:* Kinematics is studied by practitioners and researchers in different fields of practice. It is therefore critically important to adhere to a taxonomy that explicitly describes positions and movements. However, current representation methods such as cardan and Euler angles fail to report shoulder angles in a way that is easily and correctly interpreted by practitioners, and that is free from numerical instability such as gimbal lock.

*Methods:* In this paper, we comprehensively describe the recent Tilt-and-Torsion method and compare it to the Euler YXY method currently recommended by the International Society of Biomechanics. While using the same three rotations (plane of elevation, elevation, humeral rotation), the Tilt-and-Torsion method reports humeral rotation independently from the plane of elevation. We assess how it can be used to describe shoulder angles (1) in a simulated assessment of humeral rotation with the arm at the side, which constitutes a gimbal lock position, and (2) during an experimental functional task, with 10 wheelchair basketball athletes who sprint in straight line using a sports wheelchair.

*Findings:* In the simulated gimbal lock experiment, the Tilt-and-Torsion method provided both humeral elevation and rotation measurements, contrary to the Euler YXY method. During the wheelchair sprints, humeral rotation ranged from 14° (externally) to 13° (internally), which is consistent with typical maximal ranges of humeral rotation, compared to 65° to 50° with the Euler YXY method.

*Interpretation:* Based on our results, we recommend that shoulder angles be expressed using Tilt-and-Torsion angles instead of Euler YXY.

# KEYWORDS

# 1   INTRODUCTION

Kinematics of human movement is one of the main pillars of human movement biomechanics. Since kinematics is studied by practitioners and researchers who have different backgrounds ranging from engineering to medicine, it is very important to adhere to a systematic taxonomy describing positions and movements thoroughly and accurately (Doorenbosch et al., 2003; Gerhardt, 1983).

Kinematics are often described in terms of joint angles. Joint angles are well understood and unambiguous when confined to standard planes using simple projections like the Sagittal-Frontal-Transverse-Rotation (SFTR) method (Gerhardt, 1983). However, complex movements that combine multiple planes are difficult to express, especially shoulder movements due to the joint's high mobility. Over the years, many alternative conventions have been used or investigated to represent shoulder angles: attitude vector/quaternion (Woltring, 1994), cardan angles (Rab et al., 2002; van der Helm and Pronk, 1995; Wang et al., 1998), Euler angles (Davis et al., 1998; de Groot, 1997; van der Helm and Pronk, 1995; Wu et al., 2005), and the Globe system (An et al., 1991; Browne et al., 1990). All these methods seek yet somehow fail to represent shoulder angles in a way that is both (1) easily and correctly interpreted by practitioners; and (2) free of numerical instability such as gimbal lock (GL).

Recently, Campeau-Lecours et al. (2020) introduced the Tilt-and-Torsion (TT) method. This method is an extension of the Euler YXY method, the latter being recommended by the International Society of Biomechanics (ISB) (Wu et al., 2005). The TT method is believed to best describe humeral rotation while being simple to understand since it rotates the humerus around two axes instead of three. Being an extension of Euler YXY, it also has the same GL conditions. However, based on the definition of humeral rotation in TT, it is plausible that humeral rotation can still be measured even in a GL position. This needs to be verified experimentally, however.

Although TT has been tested with a single participant performing a standardized task (Campeau-Lecours et al., 2020), it has not been tested with a cohort of participants performing functional tasks. Propelling a sports wheelchair is a good example of such a task. While wheelchair propulsion has been studied extensively over the last few decades, the reported shoulder kinematics lack standardization: some authors used variations of the SFTR method (Boninger et al., 1998; Cooper et al., 1999), some used the Euler YXY method (Collinger et al., 2008; Crespo-Ruiz et al., 2011; Rao et al., 1996), while others used various cardan or Euler angle sequences (Davis et al., 1998; Lafta et al., 2018; Tsai et al., 2012) to avoid GL in a non-elevated position. Since all these methods report different angles, it is very difficult to compare results between studies. This is especially important in wheelchair sports, especially when kinematics is much different as for standard wheelchair propulsion compared to sports wheelchair propulsion.

In this paper, we describe and compare the Euler YXY and TT methods and we evaluate how they meet two generally consensual criteria: being unambiguous and being numerically stable (e.g., avoiding GL). Then, through simulation, we test if TT can express humeral rotation in a non-elevated position (GL condition). Lastly, we report and discuss the shoulder angles measured using both methods during propulsion of a sports wheelchair.

We hypothesize that (1) TT can provide clinically useful information such as humeral rotation even with a non-elevated arm; and (2) expressing shoulder kinematics during a functional task using TT yields more coherent measurements than Euler YXY.

## 2 METHODS

### 3 CARDAN AND EULER ANGLES

Cardan and Euler angles are the most popular ways to express shoulder angles. They consist of a series of three subsequent rotations, $\theta_1, \theta_2, \theta_3$, around orthogonal axes. Cardan angles are rotations along each of the three axes of a given coordinate system, while proper Euler angles have the same (local) axis for $\theta_1$ and $\theta_3$. The main difficulty associated with cardan and Euler angles is their susceptibility to a GL, which occurs when two axes become aligned in a way that prevents all three angles from being identified. For example, with Euler angles, a GL occurs when $\theta_2$ is 0° or 180°.

Given the anatomic reference position where all axes align (i.e., all angles are 0°), expressing shoulder kinematics with Euler angles causes a GL even before a movement is performed. Phadke et al. (2011) explained that in some cases, a GL in a non-elevated shoulder must be avoided, for example, when assessing the humeral rotation range of motion with the arm on the side (Rundquist et al., 2003). They therefore recommended a cardan sequence of XZY instead of Euler angles. However, cardan angles are also susceptible to a GL when the shoulder is elevated at 90°, a position that often occurs in daily activities.

In addition to the GL problem, there are 12 different combinations of cardan and Euler rotation sequences, each having a significant incidence on the returned angles (Bonnefoy-Mazure et al., 2010; Phadke et al., 2011; Šenk and Chèze, 2006). Finding the "right" sequence that returns valid angles and avoids GL is very complex, and Šenk and Chèze (2006) proposed that the rotation sequence should be selected as a function of the analyzed task. However, standardizing the way shoulder orientation is communicated is paramount. In 2005, based on work by van der Helm (1997), the ISB recommended using the Euler YXY sequence to describe shoulder motion (Wu et al., 2005). This sequence of rotations is as follows (Fig. 1a), for a right shoulder:

1. An initial rotation ($\theta_1$) along the longitudinal axis of the humerus (y). This rotation establishes the shoulder's plane of elevation. 0° relates to pure abduction, 90° to pure sagittal flexion and -90° to pure sagittal extension.
2. A second rotation ($\theta_2$) along the rotated anteroposterior axis of the humerus coordinate system (x'). This rotation corresponds to the shoulder elevation into the previously established plane of elevation. To respect geometric sign conventions, the elevation is $-\theta_2$.
3. A final rotation ($\theta_3$) along the rotated longitudinal axis of the humerus (y''). This rotation corresponds to the internal humeral rotation.

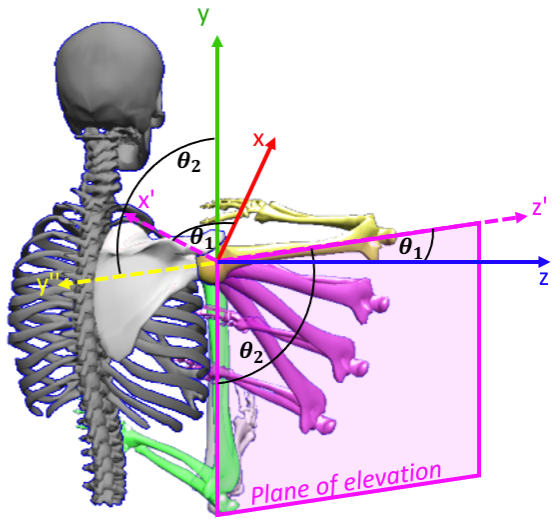

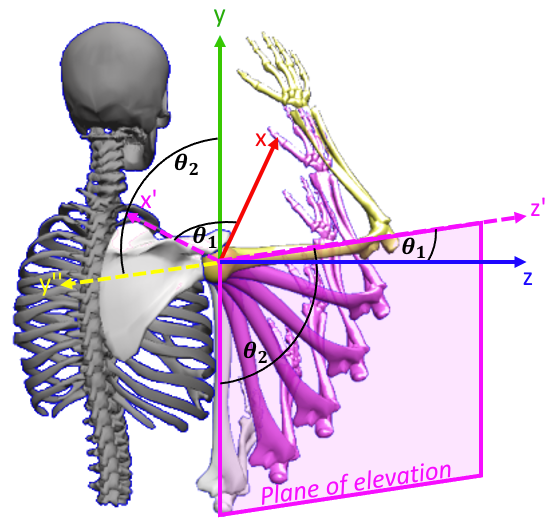

(a) Euler YXY

(b) Tilt-and-Torsion

1. The plane of elevation is established by rotating the humerus around its y axis by $\theta_1$ (green bones).

2. The humerus is elevated around its rotated x axis (x') by $\theta_2$ (here, 90°) up to its final position (yellow bones).

3. The humerus is then rotated along its final longitudinal axis (y'') by $\theta_3$ (not shown).

1. The plane of elevation is established by creating a rotation axis based on the humerus x axis, rotated around the humerus y axis by $\theta_1$.

2. The humerus is elevated directly in the plane of elevation around x' by $\theta_2$ (here, 90°), up to its final position (yellow bones).

3. The humerus is then rotated along its final longitudinal axis (y'') by $\theta_3$ (not shown).

Figure 1. Pure 90° elevation of the humerus in a 45° plane of elevation as per Euler YXY and Tilt-and-Torsion methods.

From a mathematical point of view, this method is identical to the Globe system, an alternative method that provides a way for practitioners to characterize shoulder orientation visually (Doorenbosch et al., 2003; Rab, 2008).

The Euler YXY method is well defined, and its Globe counterpart makes it easy to derive angles from a given orientation. However, interpreting these three angles may lead to unexpected orientations, particularly for movement performed in planes other than the frontal plane. This problem is illustrated for different shoulder orientations in Fig. 2a.

| Type of movement | Plane of elevation | Elevation | Humeral rotation |
|---|---|---|---|
| Abduction | 0° (frontal) | 10° to 90° | 0° |
| 90° abduction to 90° flexion | 0° to 90° | 90° | 0° |
| Flexion | 90° (sagittal) | 10° to 90° | 0° |

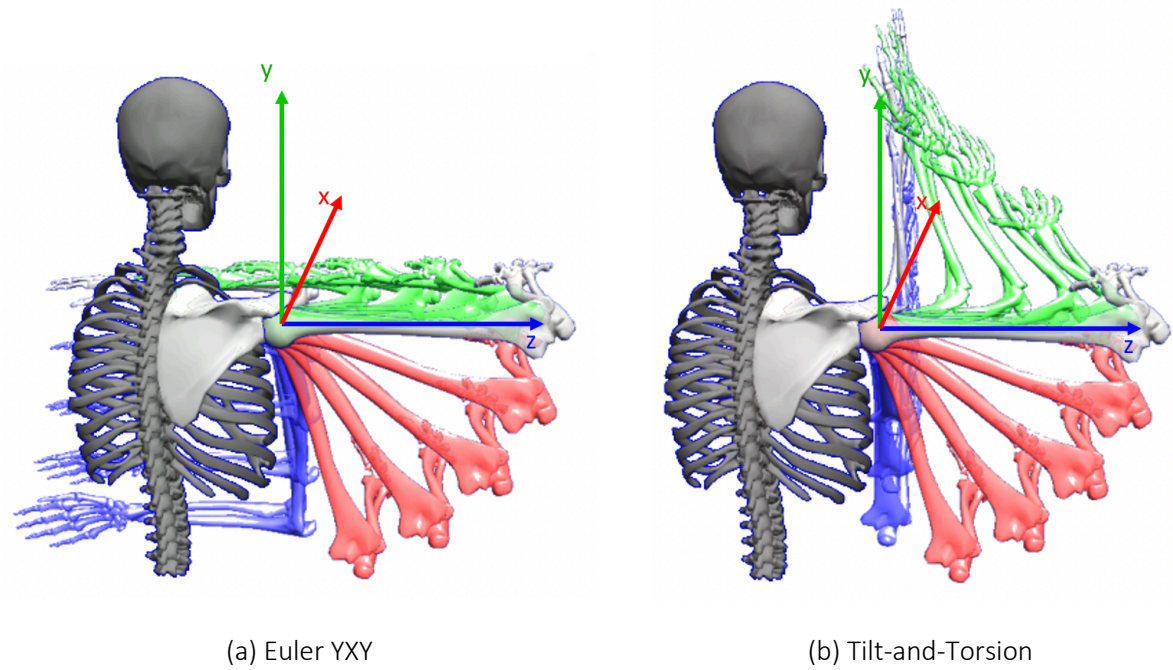

|  (a) Euler YXY | (b) Tilt-and-Torsion |
|---|---|

Figure 2. Examples of simple movements as represented by the Euler YXY and TT methods.

From the observation of Fig 2a, the following movement expressed in YXY Euler angles ($\theta_1 = 0°$; $\theta_2 \in [10°, 90°]$; $\theta_3 = 0°$) illustrated in red, describes an elevation of 10° to 90° in the frontal (0°) plane with no humeral rotation, meaning pure abduction. However, if for the same movement, $\theta_1$ was 90° (in blue), then this movement should logically refer to a pure sagittal flexion. Unfortunately, using Euler YXY, the shoulder is flexed, but also internally rotated by 90°. This bias is due to the initial rotation ($\theta_1$), which, in addition to establishing the plane of elevation, also pre-rotates the humerus. This incongruity, which was already described by Phadke et al. (2011), makes the Euler YXY method ambiguous because the reported humeral rotation is highly dependent on the plane of elevation. Unfortunately, in any combination of the three rotations, one rotation has an impact on at least one other, which means that this problem could not be solved by simply using a different rotation sequence.

## 4    TILT-AND-TORSION

The tilt-and-torsion (TT) method addresses the double humeral rotation problem by representing the shoulder orientation as a sequence of only two rotations (Campeau-Lecours et al., 2020). While still using the same three angles (plane of elevation, elevation, rotation), the humerus is rotated from its reference position directly in the plane of elevation, as shown in Fig. 1b, without being pre-rotated. The axial humeral rotation is then performed from this elevated position.

In Fig. 1b, we observe that a pure 90° elevation in a 45° plane of elevation yields a final position where the forearm is angled 45° from the horizontal plane. Although incompatible with the Euler YXY method, this is the only possible neutral humeral rotation in this plane of elevation because the rotation axis stays orthogonal to the humerus longitudinal axis throughout the entire sequence. The Euler YXY method would have incorrectly reported an external rotation of 45°.

Fig. 2b shows the same set of angles as Fig. 2a, this time based on TT. We observe that TT is consistent for all planes of elevation, not only for the frontal plane.

From an algebraic perspective, TT is only an extension of Euler YXY and is solved the same way; only the interpretation of the given angles is different:

- With Euler YXY, the plane of elevation is $\theta_1$, the elevation is $-\theta_2$ and the internal rotation is $\theta_3$.
- With TT, the plane of elevation is $\theta_1$, the elevation is $-\theta_2$ and the internal rotation is $\theta_1 + \theta_3$.

The sum of $\theta_1$ and $\theta_3$ completely makes sense since the Euler sequence rotates the humerus twice around its longitudinal axis: once for setting the plane of elevation, and a second time during the third rotation.

This new definition of humeral rotation may have a strong effect on the impact of GL on humeral rotation measurements. Campeau-Lecours et al. (2020) aptly mentioned that both Euler YXY and TT have the same GL condition. In neutral position, only $\theta_2$ and $(\theta_1 + \theta_3)$ can be solved; with 180° elevation, only $\theta_2$ and $(\theta_1 - \theta_3)$ can be solved. However, given TT's angle definitions, in the neutral position, both the elevation $(-\theta_2)$ and humeral rotation $(\theta_1 + \theta_3)$ should be solvable even within this GL condition.

We conclude this section by summarizing how to define the shoulder orientation based on three Tilt-and-Torsion angles (forward kinematics), and how to obtain the three Tilt-and-Torsion angles from a given shoulder orientation (inverse kinematics):

1) Forward kinematics: from a neutral position, the humerus is elevated by a given *elevation* angle, directly into a *plane of elevation* that is at a given angle from the frontal plane: 0° = abduction, 90° = flexion, -90° = extension, and 180° = adduction. Then, the humerus is rotated along its final longitudinal axis by a given *humeral rotation* angle.
2) Inverse kinematics: the three Euler YXY angles $\theta_1, \theta_2, \theta_3$ expressing the shoulder orientation are calculated and are interpreted using: *plane of elevation* = $\theta_1$, *elevation* = $-\theta_2$, and *humeral rotation* = $\theta_1 + \theta_3$.

To better visualize the relation between a given shoulder orientation, its Euler angles, and its TT interpretation, a visual glossary is provided in the Appendix. This glossary shows elevation and rotation movements in different standard planes of elevation.

## 5 EXPERIMENT 1: SIMULATION OF AN INTERNAL-EXTERNAL ROTATION ASSESSMENT WITH THE ARM AT THE SIDE

To verify our first hypothesis, we conducted two simulations of a pure rotation movement ranging from -100° (externally) to 100° (internally) in a non-elevated position, using Python and SciPy's spatial module. In the first simulation, the elevation was exactly 0° (i.e., pure GL), whereas in the second simulation, we added ±1° noise to the three reference angles to better reflect experimental conditions.

For both simulations, we first created the rotation matrices corresponding to this movement. Then we extracted the angles using Euler YXY and TT. With TT, we also verified that the given elevation ($-\theta_2$) and humeral rotation ($\theta_1 + \theta_3$) were identical to the reference angles, regardless of this GL condition. The code used to perform this simulation is provided as supplemental material.

## 6    EXPERIMENT 2: SHOULDER KINEMATICS DURING SPORTS WHEELCHAIR PROPULSION

To verify our second hypothesis, ten experienced wheelchair basketball athletes (Table 1) were recruited in a study assessing sports wheelchair propulsion biomechanics. To be included, participants had to be participating in competitive wheelchair basketball on a regular basis for at least one year, and have no current or recent (3 months) injury or pain. The experimental protocol was approved by the Institutional Research Ethics Committee of Université du Québec à Montréal (UQAM) (certificate #CIEREH 2879_e_2018).

All participants performed four, 9-m sprints at maximal speed in a straight line from a stopped position on a wooden basketball court. They were asked to propel synchronously (both arms pushing at the same time). The participants' wheelchairs were equipped with two synchronized instrumented wheels (SmartWheel) that measured the propulsion forces. An optoelectronic system consisting of 14 cameras (Prime13, Optitrack) was used to measure the participants' and wheelchair kinematics unilaterally throughout each sprint. The position of the following landmarks were recorded at 120 Hz: ulnar styloid, lateral and medial elbow epicondyles, acromion, C7, T12, and both rear wheel centers. Landmarks that could not be tracked directly due to occlusion (e.g., rear wheel center of the opposite side, medial elbow epicondyle) were reconstructed using rigid bodies composed of three to four markers affixed on the wheelchair and arms. The participants propelled twice in both directions, allowing kinematics from both sides to be recorded. All trials were processed as the right side, therefore markers on the left side were mirrored through the wheelchair's median plane.

The humerus coordinate system was defined by following ISB recommendations (Wu et al., 2005) using both elbow epicondyles and the styloid process, and approximating the glenohumeral joint based on the acromion. The trunk could only be defined by markers on participants' back (T12 and C7) because many participants had a strong forward-inclined posture that would not allow markers on the chest to be tracked. Therefore, the trunk coordinate system was built on the assumption that it was not axially rotated, which is realistic since the participants were instructed to propel synchronously. The trunk's y axis was the line from T12 to C7 and its x axis was perpendicular to y and the wheelchair's mediolateral axis, pointing forward, which is consistent with the ISB recommendations for the local thorax coordinate system.

## Table 1. Participants' anthropometrics

| Participant | Gender | Age (years) | Dominant limb | Disorder | Height (m) | Weight (kg) | BMI (kg/m²) | Experience (years) |
|---|---|---|---|---|---|---|---|---|
| 1 | F | 31 | R | Spinal cord injury (T6, A) | 1.60 | 61 | 23.8 | 3 |
| 2 | M | 60 | R | Spinal cord injury (D6-D7, A) | 1.83 | 71 | 21.2 | 6 |
| 3 | M | 29 | L | Cerebral palsy | 1.68 | 60 | 21.3 | 10 |
| 4 | M | 40 | R | Spinal cord injury (T12, A) | 1.75 | 66 | 21.6 | 12 |
| 5 | M | 34 | R | Spinal cord injury (T7) | 1.50 | 73 | 32.4 | 10 |
| 6 | M | 33 | R | Spinal cord injury (T10, A) | 1.76 | 95 | 30.7 | 1.5 |
| 7 | M | 32 | R | Muscular dystrophy | 1.73 | 52 | 17.4 | 6 |
| 8 | M | 23 | R | Spastic dysplasia | 1.63 | 58 | 21.8 | 11 |
| 9 | M | 24 | R | None (non-disabled) | 1.78 | 78 | 24.6 | 16 |
| 10 | F | 30 | R | None (non-disabled) | 1.61 | 62 | 23.9 | 3 |
| Mean | M = 8 | 33.6 | R = 9 | / | 1.69 | 67.6 | 23.9 | 7.9 |
| (SD) | F = 2 | (10.5) | L = 1 | | (0.10) | (12.3) | (4.5) | (4.7) |

Shoulder angles were calculated using both Euler YXY and TT, then separated into pushes and recoveries using a manual inspection of the propulsion forces. Pushes 1 and 2 of every sprint were excluded since they were considered transitional. The following descriptive variables were calculated for each push and recovery: minimal, mean, maximal values and range for the plane of elevation, elevation, and internal rotation. These variables were averaged for each participant and were reported as the inter-participant mean and standard deviation.

Additionally, the profile of the three angles were plotted for both methods and every participant, after being time-normalized to the push and recovery phases.

# 7 Results

## 8 Experiment 1: Simulation of an internal-external rotation assessment with the arm at the side

Fig. 3 shows the three simulated angles as given by both methods. For the simulation without noise (Fig. 3a), SciPy expectedly warned that a GL was detected, that it could not determine all the angles and that it adjusted the third angle to zero. Therefore, with Euler YXY, both the first and third angles are wrong. However, TT was able to provide the correct humeral rotation angle.

For the simulation with added noise (Fig. 3b), the noise prevented SciPy from detecting a GL. The three angles were therefore reported without a warning, but Euler YXY's first and third angles are still wrong. TT was still able to restore both the elevation and humeral rotation.

In both simulations, there was no difference in elevation or internal rotation between the reference angles and those reported by the TT method.

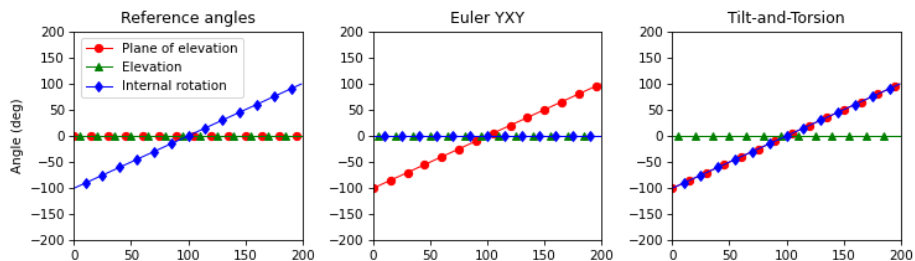

(a) Simulation in pure 0° elevation.

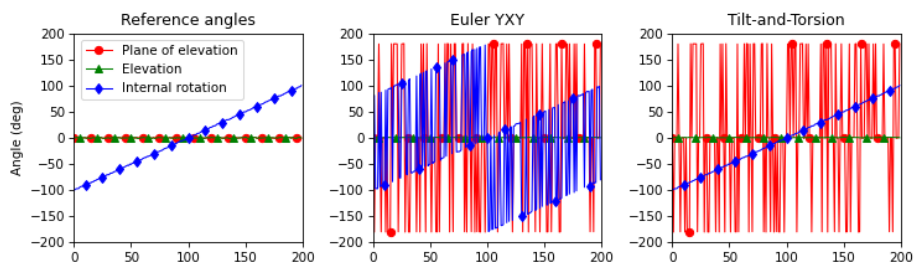

(b) Simulation with 0° elevation plus ±1° noise added to every reference angle.

Figure 3. Results of a simulated assessment of humeral rotation with arm at the side

## 9 Experiment 2: Shoulder kinematics in sports wheelchair propulsion

The descriptive values for sports wheelchair propulsion are provided in Table 2. While the plane of elevation and elevation were the same for both methods, the internal rotation was much different, ranging between 116° and 123° respectively during the push and recovery phases as reported by Euler

YXY, compared to only 22° and 26° as reported by TT. The minimal and maximal humeral rotation values were also much different between methods, with a maximal internal rotation of 58° (Euler YXY) compared to 13° (TT) and a maximal external rotation of 65° (Euler YXY) compared to 14° (TT).

The angle profiles are illustrated in Fig. 4. In the case of Euler YXY, the association between the plane of elevation and humeral rotation is obvious in Fig.s 4a and 4c. The humeral rotation obtained by the TT method, as shown in Fig. 4d, was much less variable, remaining in a ±25° range.

Table 2. Shoulder angles during propulsion of a sports wheelchair, in degrees

| | | Euler YXY | | Tilt-and-Torsion | |
|---|---|---|---|---|---|
| Push phase | | | | | |
| Plane of elevation | min | | -50 (16) | | |
| | mean | | -6 (15) | | |
| | max | | 52 (12) | | |
| | range | | 102 (16) | | |
| Elevation | min | | 28 (5) | | |
| | mean | | 37 (7) | | |
| | max | | 53 (8) | | |
| | range | | 25 (7) | | |
| Internal rotation | min | -59 | (25) | -9 | (16) |
| | mean | 12 | (20) | 5 | (12) |
| | max | 57 | (13) | 13 | (12) |
| | range | 116 | (18) | 22 | (10) |
| Recovery phase | | | | | |
| Plane of elevation | min | | -50 (13) | | |
| | mean | | 5 (17) | | |
| | max | | 53 (14) | | |
| | range | | 104 (15) | | |
| Elevation | min | | 30 (6) | | |
| | mean | | 41 (6) | | |
| | max | | 57 (9) | | |
| | range | | 27 (11) | | |
| Internal rotation | min | -65 | (25) | -14 | (16) |
| | mean | -3 | (21) | 2 | (10) |
| | max | 58 | (12) | 12 | (11) |
| | range | 123 | (17) | 26 | (14) |

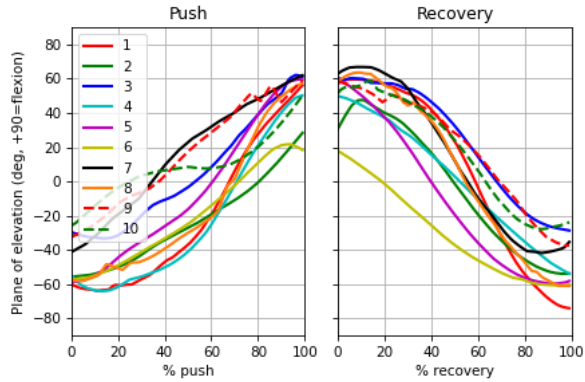

(a) Plane of elevation (Euler YXY, TT)

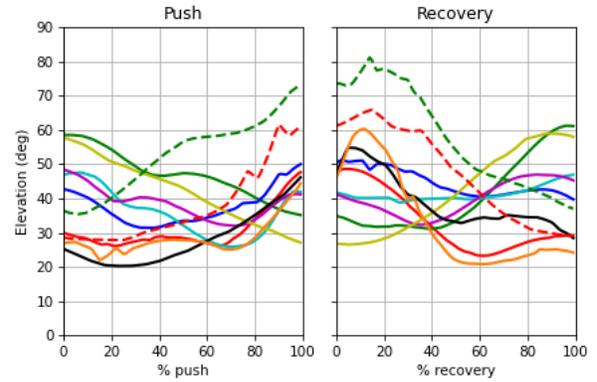

(b) Elevation (Euler YXY, TT)

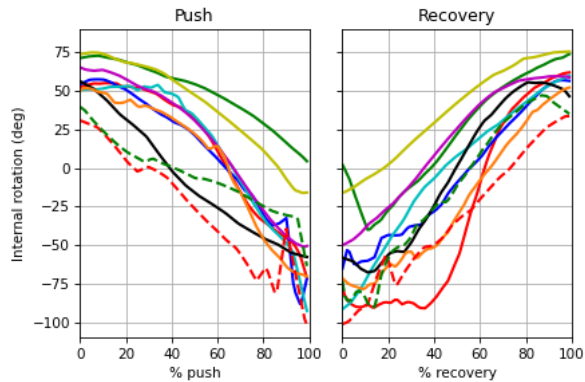

(c) Internal rotation (Euler YXY)

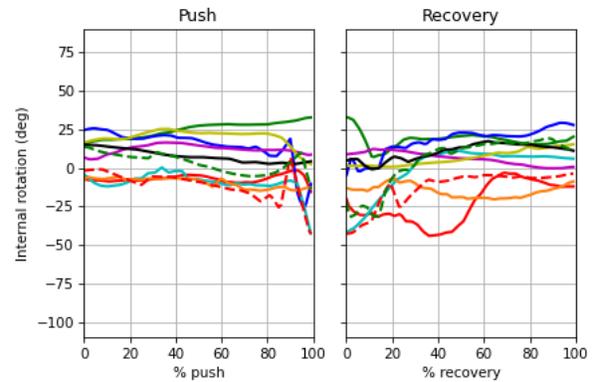

(d) Internal rotation (TT)

Figure 4. Shoulder angles during propulsion of a sports wheelchair. One cycle per participant (with the one deemed the most representative being shown).

# 10 Discussion

Our first hypothesis states that the TT method would provide valuable clinical information even with the arm at the side, which is a GL condition. This was confirmed via simulation in our first experiment. With TT, the only impact of the GL was not being able to obtain the plane of elevation. However, a plane of elevation could not be obtained if the shoulder was not elevated; thus, this GL condition does not prevent clinical insights on the analyzed movement from being gained. It is worth noting that this would not be valid in the other GL position (180° elevation) because the returned angles would be $\theta_2$ and $\theta_1 - \theta_3$ instead of $\theta_2$ and $\theta_1 + \theta_3$.

Among other authors, Šenk and Chèze (2006) considered the absence of a GL as the primary criteria for defining the best method to express shoulder angles. Since it is impossible to avoid a GL in any possible movement, then the best strategy may be to allow a GL in positions where its impact has the less clinical meaning. This is what the TT method offers.

Our second hypothesis was that TT would give more coherent angles than Euler YXY. In the second experiment, internal humeral rotation was much different between both methods, with a maximum of 58° using Euler YXY versus about 13° using TT. Other authors also reported a high maximal internal rotation value using Euler YXY: Rao et al. (1996) reported 86° for 16 persons with paraplegia who propelled a standard wheelchair on an ergometer. In a similar setup with 61 persons with paraplegia, Collinger et al. (2008) reported 84°. In an experiment with 10 wheelchair basketball players who propelled their own sport wheelchair on a treadmill, Crespo-Ruiz et al. (2011) reported 63°. Regardless of the population, these values are very large and can be considered implausible, since the maximal internal humeral rotation is approximately 50° when the arm is abducted (Barnes et al., 2001). These high values are undoubtedly related to the coupling between the plane of elevation and the humeral rotation, which is clearly visible in Fig. 4a and 4c, and in similar figures as reported by other authors (Collinger et al., 2008; Rao et al., 1996). The graphical glossary in the Appendix shows the tendency of the third Euler angle to have a positive bias for an extended shoulder, and a negative bias for a flexed shoulder. Therefore, interpreting the third Euler angle as the humeral rotation may generate false conclusions, such as exaggerating the risk of musculoskeletal disorders that would be wrongly attributed to high humeral rotation. Since Euler YXY and the Globe system are equivalent (Doorenbosch et al., 2003; Rab, 2008), then this also applies to the Globe system.

We remember that with TT, the shoulder elevation is free of humeral rotation because it is performed around an axis that is always normal to the humeral longitudinal axis. In our work, the reported values for humeral rotation using TT were much lower and also more constant, with a range of 26° instead of 123° for Euler YXY. This is a very important difference in the kinematic analysis of this movement since the humeral rotation range goes from the most important of the three angles (using Euler YXY) to the least important (using TT).

The presented study does have a few limitations. In the first experiment, the assessment of internal/external rotation was only based on a simulation to verify if the calculation of the angles was not only theoretically, but also algorithmically correct, even in a case of GL. A real assessment of shoulder rotation range of motion should be conducted to confirm these results experimentally.

In terms of experimental acquisitions of wheelchair propulsion, the assumption of a lack of trunk rotation may have slightly altered the calculation of the plane of elevation. However, since propulsion was synchronous, we believe that its impact is negligible, particularly given the very large differences in humeral rotation between both methods.

# 11  Conclusion

In this paper, we compared the Tilt-and-Torsion method to the ISB-recommended Euler YXY method to express shoulder angles that are both unambiguous and numerically stable. We found that the TT method allows clinicians to obtain insightful measurements even with the arm at the side (one of the two GL positions). We also found that the humeral rotation given by the TT method is expectedly more coherent than the Euler YXY method, and that this difference is quite significant for functional movements in other planes than the frontal plane, such as during sports wheelchair propulsion. Based on our results, we recommend that shoulder angles be expressed using TT instead of Euler YXY.

## 12 CONFLICT OF INTEREST STATEMENT

The authors do not report any conflicts of interest. The authors alone are responsible for the written content of this article.

## 13 ACKNOWLEDGMENTS

The authors want to thank Étienne Marquis, Guy El Hajj Boutros, Kim Lefebvre, and Pierre-Olivier Bédard for helping with the data acquisitions process, and Pierre-Emmanuel Porquet, Marine Gailhard and the athletes of the Centre d'intégration à la vie active (CIVA), Montreal, for their participation in this project. Data collection was funded by the Fonds de recherche du Québec via the Société Inclusive intersectoral initiative.

## 14 REFERENCES

An, K.-N., Browne, A.O., Korinek, S., Tanaka, S., Morrey, B.F., 1991. Three-dimensional kinematics of glenohumeral elevation. J. Orthop. Res. 9, 143–149. https://doi.org/10.1002/jor.1100090117

Barnes, C.J., Van Steyn, S.J., Fischer, R.A., 2001. The effects of age, sex, and shoulder dominance on range of motion of the shoulder. Journal of Shoulder and Elbow Surgery 10, 242–246. https://doi.org/10.1067/mse.2001.115270

Boninger, M.L., Cooper, R. a, Shimada, S.D., Rudy, T.E., 1998. Shoulder and elbow motion during two speeds of wheelchair propulsion: a description using a local coordinate system. Spinal cord : the official journal of the International Medical Society of Paraplegia 36, 418–426. https://doi.org/10.1038/sj.sc.3100588

Bonnefoy-Mazure, A., Slawinski, J., Riquet, A., Lévèque, J.-M., Miller, C., Chèze, L., 2010. Rotation sequence is an important factor in shoulder kinematics. Application to the elite players' flat serves. Journal of Biomechanics 43, 2022–2025. https://doi.org/10.1016/j.jbiomech.2010.03.028

Browne, A., Hoffmeyer, P., Tanaka, S., An, K., Morrey, B., 1990. Glenohumeral elevation studied in three dimensions. The Journal of Bone and Joint Surgery. British volume 72-B, 843–845. https://doi.org/10.1302/0301-620X.72B5.2211768

Campeau-Lecours, A., Vu, D.-S., Schweitzer, F., Roy, J.-S., 2020. Alternative Representation of the Shoulder Orientation Based on the Tilt-and-Torsion Angles. Journal of Biomechanical Engineering 142, 074504. https://doi.org/10.1115/1.4046203

Collinger, J.L., Boninger, M.L., Koontz, A.M., Price, R., Sisto, S.A., Tolerico, M.L., Cooper, R.A., 2008. Shoulder biomechanics during the push phase of wheelchair propulsion: a multisite study of persons with paraplegia. Arch Phys Med Rehabil 89, 667–676.

Cooper, R.A., Boninger, M.L., Shimada, S.D., Lawrence, B.M., 1999. Glenohumeral joint kinematics and kinetics for three coordinate system representations during wheelchair propulsion. American journal of physical medicine & rehabilitation 78, 435–446.

Crespo-Ruiz, B., Ama-Espinosa, A. del, Gil-Agudo, Á., 2011. Relation between kinematic analysis of wheelchair propulsion and wheelchair functional basketball classification. Adapted physical activity quarterly : APAQ 28, 157–172. https://doi.org/10.1123/apaq.28.2.157

Davis, J.L., Growney, E.S., Johnson, M.E., Iuliano, B.A., An, K.N., 1998. Three-dimensional kinematics of the shoulder complex during wheelchair propulsion: A technical report. Bulletin of prosthetics research 35, 61–72.

de Groot, J.H., 1997. The variability of shoulder motions recorded by means of palpation. Clinical Biomechanics 12, 461–472. https://doi.org/10.1016/S0268-0033(97)00031-4

Doorenbosch, C.A.M., Harlaar, J., Veeger, D. (H. E.J.), 2003. The globe system: An unambiguous description of shoulder positions in daily life movements. The Journal of Rehabilitation Research and Development 40, 149. https://doi.org/10.1682/JRRD.2003.03.0149

Gerhardt, J.J., 1983. Clinical measurements of joint motion and position in the neutral-zero method and SFTR recording: Basic principles. International Rehabilitation Medicine 5, 161–164. https://doi.org/10.3109/03790798309167039

Lafta, H.A., Guppy, R., Whatling, G., Holt, C., 2018. Impact of rear wheel axle position on upper limb kinematics and electromyography during manual wheelchair use. International Biomechanics 5, 17–29. https://doi.org/10.1080/23335432.2018.1457983

Phadke, V., Braman, J.P., LaPrade, R.F., Ludewig, P.M., 2011. Comparison of glenohumeral motion using different rotation sequences. Journal of Biomechanics 44, 700–705. https://doi.org/10.1016/j.jbiomech.2010.10.042

Rab, G., Petuskey, K., Bagley, A., 2002. A method for determination of upper extremity kinematics. Gait & Posture 15, 113–119. https://doi.org/10.1016/S0966-6362(01)00155-2

Rab, G.T., 2008. Shoulder motion description: The ISB and Globe methods are identical. Gait & Posture 27, 702–705. https://doi.org/10.1016/j.gaitpost.2007.07.003

Rao, S.S., Bontrager, E.L., Gronley, J.K., Newsam, C.J., Perry, J., 1996. Three-dimensional kinematics of wheelchair propulsion. IEEE Trans Rehabil Eng 4, 152–160.

Rundquist, P.J., Anderson, D.D., Guanche, C.A., Ludewig, P.M., 2003. Shoulder Kinematics in Subjects With Frozen Shoulder 84, 7.

Šenk, M., Chèze, L., 2006. Rotation sequence as an important factor in shoulder kinematics. Clinical Biomechanics 21, S3–S8. https://doi.org/10.1016/j.clinbiomech.2005.09.007

Tsai, C.-Y., Lin, C.-J., Huang, Y.-C., Lin, P.-C., Su, F.-C., 2012. The effects of rear-wheel camber on the kinematics of upper extremity during wheelchair propulsion. BioMedical Engineering OnLine 11, 87. https://doi.org/10.1186/1475-925X-11-87

van der Helm, F.C., 1997. A standardized protocol for motion recordings of the shoulder, in: First Conference of the International Shoulder Group. Shaker Publishers BV, Delft University of Technology The Netherlands, pp. 7–12.

van der Helm, F.C.T., Pronk, G.M., 1995. Three-Dimensional Recording and Description of Motions of the Shoulder Mechanism. Journal of Biomechanical Engineering 117, 27–40. https://doi.org/10.1115/1.2792267

Wang, X., Maurin, M., Mazet, F., Maia, N.D.C., Voinot, K., Verriest, J.P., Fayet, M., 1998. Three-dimensional modelling of the motion range of axial rotation of the upper arm. Journal of Biomechanics 31, 899–908. https://doi.org/10.1016/S0021-9290(98)00098-0

Woltring, H.J., 1994. 3-d attitude representation of human joints: a standardization proposal. J. Biomechanics 27, 1399–1414.

Wu, G., Van Der Helm, F.C.T., Veeger, H.E.J.D., Makhsous, M., Van Roy, P., Anglin, C., Nagels, J., Karduna, A.R., McQuade, K., Wang, X., Werner, F.W., Buchholz, B., Others, 2005. ISB recommendation on definitions of joint coordinate systems of various joints for the reporting of human joint motion - Part II: shoulder, elbow, wrist and hand. Journal of Biomechanics 38, 981–992. https://doi.org/10.1016/j.jbiomech.2004.05.042

Appendix: Graphical glossary

**Top section (rows: Euler YXY, TT)**

| | Neutral position | Posterior elevation (extension) | Lateral elevation (abduction) | Elevation in the scapular plane (scaption) | Anterior elevation (flexion) |
|---|---|---|---|---|---|
| *Euler YXY* $\theta_1$ (y) | undefined | -90° | 0° | 30° | 90° |
| $\theta_2$ (x') | 0° | 0° to -90° | 0° to -90° | 0° to -90° | 0° to -90° |
| $\theta_3$ (y'') | undefined | 90° | 0° | -30° | -90° |
| **Plane of elev. ($\theta_1$)** | undefined | -90° | 0° | 30° | 90° |
| **Elevation ($-\theta_2$)** | 0° | 0° to 90° | 0° to 90° | 0° to 90° | 0° to 90° |
| **Internal rotation ($\theta_1 + \theta_3$)** | 0° | 0° | 0° | 0° | 0° |

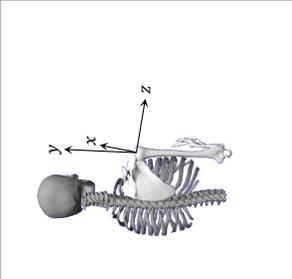 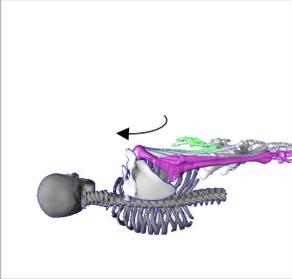 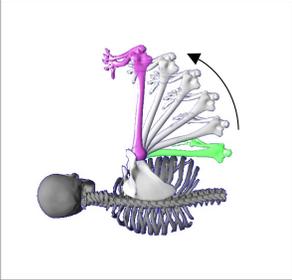 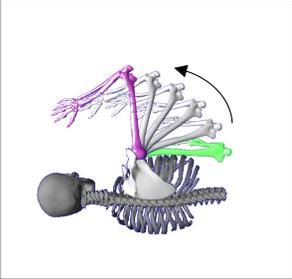

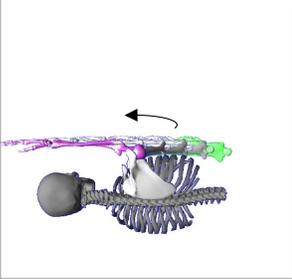 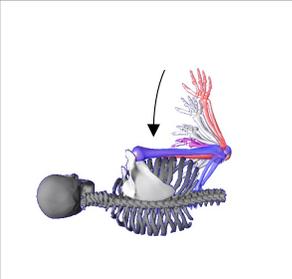 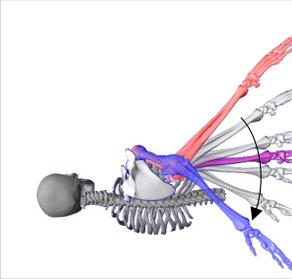 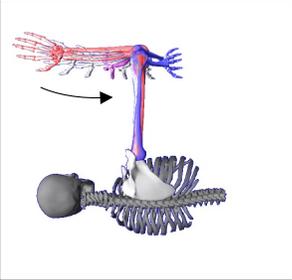

**Bottom section (rows: Euler YXY, TT)**

| | Internal rotation in a non-elevated position | Internal rotation in an extended position | Internal rotation in an abducted position | Internal rotation in an elevated position in the scapular plane | Internal rotation in a flexed position |
|---|---|---|---|---|---|
| *Euler YXY* $\theta_1$ (y) | undefined | -90° | 0° | 30° | 90° |
| $\theta_2$ (x') | 0° | -90° | -90° | -90° | -90° |
| $\theta_3$ (y'') | undefined | 45° to 90° to 135° | -45° to 0° to 45° | -75° to -30° to 15° | -135° to -90° to -45° |
| **Plane of elevation** | undefined | -90° | 0° | 30° | 90° |
| **Elevation** | 0° | 90° | 90° | 90° | 90° |
| **Int. rotation** | -45° to 0° to 45° | -45° to 0° to 45° | -45° to 0° to 45° | -45° to 0° to 45° | -45° to 0° to 45° |

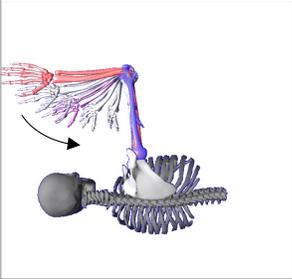 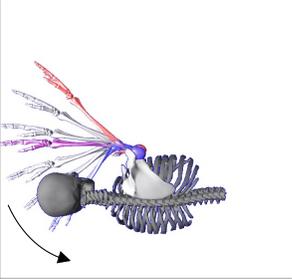 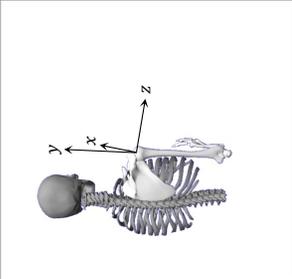 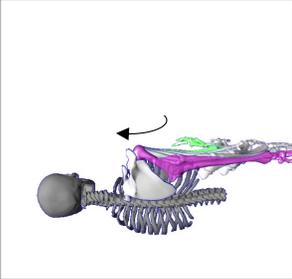 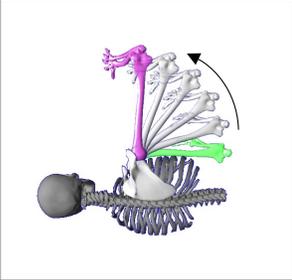

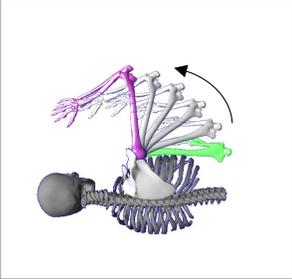 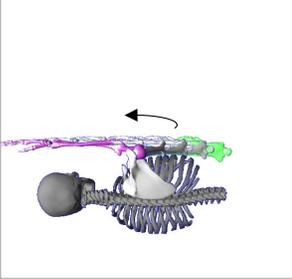 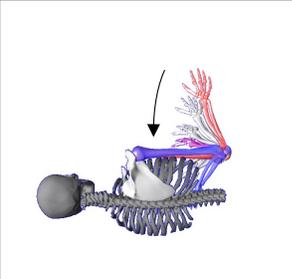 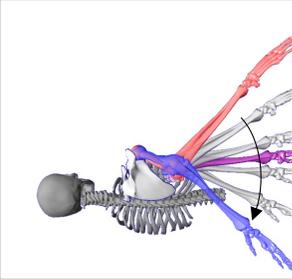 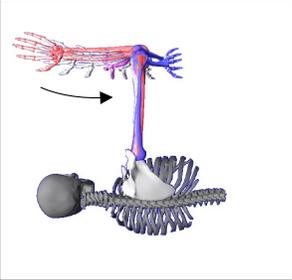

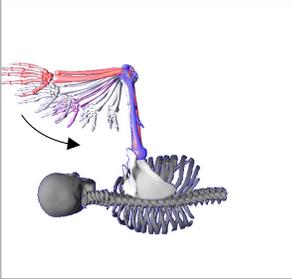 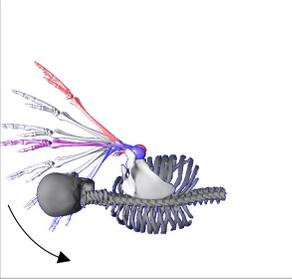

Supplementary material

```python
"""
Simulation of shoulder rotation assessment with arm at the side.

The following code was used to simulate the internal-external humeral rotation assessment
in a gimbal lock position and to generate Figure 3.

"""

import scipy.spatial.transform as transform
import numpy as np
import matplotlib.pyplot as plt

def yxy_to_tt(yxy_angles):
    """Convert YXY angles to Tilt-and-Torsion angles."""
    tt_angles = yxy_angles.copy()
    tt_angles[:, 2] += tt_angles[:, 0]
    tt_angles = np.mod(tt_angles + 180, 360) - 180
    return tt_angles

def tt_to_yxy(tt_angles):
    """Convert Tilt-and-Torsion angles to YXY angles."""
    yxy_angles = tt_angles.copy()
    yxy_angles[:, 2] -= yxy_angles[:, 0]
    yxy_angles = np.mod(yxy_angles + 180, 360) - 180
    return yxy_angles

def plot_angles(angles, title, i_plot):
    """Plot angles, set axes and write title."""
    plt.subplot(1, 3, i_plot)
    plt.plot(angles[:, 0], 'ro-', markevery=(0, 15), label='Plane of elevation')
    plt.plot(angles[:, 1], 'g^-', markevery=(5, 15), label='Elevation')
    plt.plot(angles[:, 2], 'bd-', markevery=(10, 15), label='Internal rotation')
    plt.axis([0, 200, -200, 200])
    plt.title(title)

    if i_plot == 1:
        plt.ylabel('Angle (deg)')
        plt.legend()

# Running conditions
add_noise = True  # False for Fig. a, True for Fig. b
np.random.seed(0)

# Define a movement of pure humeral rotation, arm at the side.
reference_angles = np.empty((200, 3))
reference_angles[:, 0] = 0  # Plane of elevation = 0
reference_angles[:, 1] = 0  # Elevation = 0
reference_angles[:, 2] = np.arange(-100, 100)  # Rotation of -100 to 100 deg

# Add white noise of ±1 amplitude
if add_noise:
    reference_angles += 2 * (np.random.rand(200, 3) - 0.5)
```

```python
# Convert the reference angles to a series of rotation matrices
matrices = transform.Rotation.from_euler(
    'YXY', tt_to_yxy(reference_angles), degrees=True)

# Extract the angles back using YXY and Tilt-Torsion methods
yxy_extracted_angles = matrices.as_euler('YXY', degrees=True)
tt_extracted_angles = yxy_to_tt(yxy_extracted_angles)

# Plot the results
plot_angles(reference_angles, 'Reference angles', 1)
plot_angles(yxy_extracted_angles, 'Euler YXY', 2)
plot_angles(tt_extracted_angles, 'Tilt-and-Torsion', 3)

# Compare the internal rotation between reference and tt_extracted
assert np.allclose(tt_extracted_angles[:, 2], reference_angles[:, 2])
```